\documentclass
[preprintnumbers,showkeys,showpacs,superscriptaddress,nofootinbib,byrevtex]{revtex4}
\usepackage{amsmath,amsfonts,amssymb,amscd,amsxtra,amsthm}
\usepackage{graphicx}
\usepackage{bm}
\begin{document}
\preprint{INHA-NTG-08/2011}
\title{$\Lambda(1405,1/2^-)$ photoproduction from the $\gamma p\to 
  K^+\Lambda(1405)$ reaction} 
\author{Seung-il Nam}
\email[E-mail: ]{sinam@kau.ac.kr}
\affiliation{Research Institute of Basic Sciences, Korea Aerospace
University, Goyang, 412-791, Korea} 
\author{Ji-Hyoung~Park}
\email[E-mail: ]{parkjihyoung@pusan.ac.kr}
\affiliation{Department of Physics,
Pusan National University, Busan 609-735, Republic of Korea} 
\author{Atsushi Hosaka}
\email[E-mail: ]{hosaka@rcnp.osaka-u.ac.jp}
\affiliation{Research Center for Nuclear Physics (RCNP), Ibaraki,
  Osaka 
567-0047, Japan}
\author{Hyun-Chul~Kim}
\email[E-mail: ]{hchkim@inha.ac.kr}
\affiliation{Department of Physics, Inha University, 
Incheon 402-751, Republic of Korea}  
\date{June, 2011}
\begin{abstract} 
We investigate the photoproduction of
$\Lambda(1405,1/2^-)\equiv\Lambda^*$ off the proton target using the  
effective Lagrangian in the Born approximation. We observed that,
depending on the choice of the $K^*N\Lambda^*$ coupling 
strength, the total cross section becomes
$0.1\lesssim\sigma_{\Lambda^*}\lesssim0.2\,\mu\mathrm{b}$ near the 
threshold and starts to decrease beyond $E_\gamma\approx 1.6$ GeV, and
the angular dependence shows a mild enhancement in the forward
direction. It turns out that the energy dependence of the total cross
section is similar to that shown in the recent LEPS
experiment. This suggests that the production mechanism of the
$\Lambda^*$ is dominated by the $s$-channel contribution. 
\end{abstract} 
\maketitle
\section{Introduction}
The $\Lambda(1405,1/2^-)\equiv\Lambda^*$, an excited state of the 
$\Lambda(1116,1/2^+)$ with negative parity, has been studied for
decades (See Ref.~\cite{Dalitz:1998jx} for the review until 1998).
Its production has been mainly conducted in proton-proton 
scattering and meson-proton scattering.  Recently, however, the LEPS
collaboration has carried out the measurement of the
$\Lambda^*$ produced in the $\gamma p\to K^+
\Lambda^*$ reaction~\cite{Fujimura:2007zz,Niiyama:2008rt}.     
While for the ground state $\Lambda(1116,1/2^+)$ and for the $d$-wave 
$\Lambda(1520,3/2^-)$ photoproductions, there are many 
experimental data~\cite{Tran:1998qw,Barber:1980zv} as well as
theoretical investigations~\cite{Janssen:2001pe,Janssen:2001wk,
Nam:2005uq,Ozaki:2007ka,Roca:2004wt}, there has been only a few
theoretical works on the $\Lambda^*$ photoproduction to 
date~\cite{Williams:1991tw,Nacher:1998mi,Lutz:2004sg} and 
related experiments mainly performed by the LEPS 
collaboration at SPring-8~\cite{Fujimura:2007zz,Niiyama:2008rt}.

Nevertheless, there are several interesting theoretical
works. For example, Ref.~\cite{Williams:1991tw} estimated the  
differential cross section for the $\gamma p\to K^+\Lambda^*$
reaction, considering the crossing symmetry and duality, 
whereas Refs.~\cite{Nacher:1998mi,Lutz:2004sg} concentrated on the
$\Lambda^*$ invariant mass spectrum via the $\gamma p\to K^+\pi\Sigma$
scattering process using the $s$-wave chiral dynamics, also known as
the chiral unitary model ($\chi$UM), in which the $\Lambda^*$ is
assumed to be the molecular-type $\bar{K}N$ state rather than a
three-quark color-singlet ($uds$) one such as usual baryons. In the
recent LEPS experiment~\cite{Niiyama:2008rt}, interestingly, it turned
out that the $\Lambda^*/\Sigma(1385)$ production ratio is very
different between the low ($1.5\lesssim E_\gamma\lesssim2.0$ GeV) and
high ($2.0\lesssim E_\gamma\lesssim2.4$ GeV) energy regions. It
implies that the total cross section for the $\Lambda^*$ increases
near the threshold, and then starts to decrease as the photon 
energy is increased. It was suggested that this interesting tendency
may be caused by either the different production mechanisms from that
for the $\Sigma(1385)$ or the novel internal structure of the
$\Lambda^*$.    

In the present work, we aim at investigating the $\gamma
p\to K^+\Lambda^*$ reaction, using the effective Lagrangian
in the Born approximation. We make use of theoretical and experimental
information to determine relevant parameters such as the 
coupling strengths and cutoff masses for the phenomenological form
factors, which are treated in a gauge-invariant manner. By changing
the cutoff mass for the phenomenological form factor for the
$\gamma\Lambda^*\Lambda^*$ vertex, the size effect, which may encode
the internal structure of the $\Lambda^*$, is examined.

We observe that, depending on the choice of the $K^*N\Lambda^*$
coupling strength, the total cross section becomes
$0.1\lesssim\sigma_{\Lambda^*}\lesssim0.2\,\mu\mathrm{b}$ near the
threshold and starts to decrease slowly beyond $E_\gamma\approx1.6$
GeV, and the angular dependence shows a mild enhancement in the
forward direction.  It is also found that the size effect of the
$\Lambda^*$ is seen mainly due to the $u$-channel near the threshold but very
small. Comparing these results to the experimental
data~\cite{Niiyama:2008rt}, the overall energy dependence of the 
$\sigma_{\Lambda^*}$ is very similar. This indicates that the
production mechanism of the $\Lambda^*$ is dominated by the
$s$-channel contribution rather than the $t$-channel 
one. The photon-beam asymmetry shows a strong electric photon-hadron
coupling contribution due to the $t$-channel.     

We organize the present work as follows: In Section II, we briefly  
explain the general formalism relevant for studying the $\gamma p\to
K^+\Lambda^*$ scattering process.  In Section III, the numerical
results are given with discussions.  Theoretical ambiguities are
briefly explained in Section IV. The final Section is devoted to
summarize the present work and to draw conclusions.   

\section{General Formalism\label{sec:2}}
\begin{figure}[t]
\includegraphics[width=12cm]{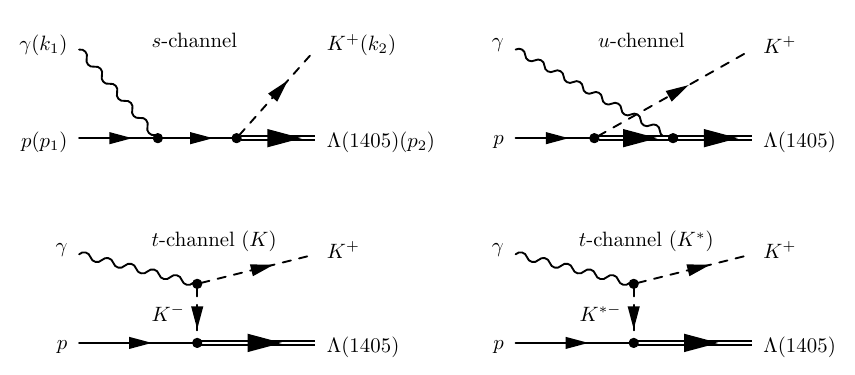}
\caption{Fyenman diagrams for the $\Lambda^*$ photoproduction from the
  proton target, $\gamma p\to K^+\Lambda(1405)$, in the pseudoscalar
  (PS) meson-baryon coupling scheme.}       
\label{fig0}
\end{figure}
We start with the general formalism for the
$\Lambda(1405,1/2^-)\equiv\Lambda^*$ photoproduction. 
In Fig.~\ref{fig0}, we depict the relevant Feynman diagrams.  The four 
momenta for the involved particles are defined in the figure. To
compute the diagrams, we employ the following effective Lagrangian
in the pseudoscalar (PS) meson-baryon coupling scheme: 
\begin{eqnarray}
\label{eq:EL1405}
\mathcal{L}_{\gamma KK}
&=&ie_K\left[
(\partial^{\mu}K^{\dagger})K-(\partial^{\mu}K)K^{\dagger}
\right]A_{\mu}+ {\rm h.c.},
\nonumber\\
\mathcal{L}_{\gamma{NN}}
&=&
-\bar{N}\left[e_N\rlap{/}{A}
+\frac{\kappa_{N}}{2M_N}
\sigma_{\mu\nu}F^{\mu\nu}\right]N+ {\rm h.c.},
\nonumber\\
\mathcal{L}_{\gamma\Lambda^*\Lambda^*}
&=&
-\frac{\kappa_{\Lambda^*}}{2M_{\Lambda^*}}
\bar{\Lambda}^*\sigma_{\mu\nu}F^{\mu\nu}\Lambda^*+ {\rm h.c.},
\nonumber\\
\mathcal{L}_{KN\Lambda^*}
&=&ig_{KN\Lambda^*}
\bar{\Lambda}^*K^{\dagger}N+ {\rm h.c.},
\nonumber\\
\mathcal{L}_{\gamma {K} K^{*}}&=& g_{\gamma K K^{*}}
\epsilon_{\mu\nu\sigma\rho}(\partial^{\mu}A^{\nu})
(\partial^{\sigma}K^{\dagger}){K}^{*\rho} 
+ {\rm h.c.},
\nonumber\\
\label{gkv}
\mathcal{L}_{K^{*}N\Lambda^*}&=&g_{K^{*}N\Lambda^*}\bar{\Lambda}^*
\gamma^{\mu}\gamma_{5}{K}^{*\dagger}_{\mu}N
+ {\rm h.c.}, 
\end{eqnarray}
where $K$, $A_{\mu}$, $N$, $\Lambda^*$, and $K^*$ represent the
pseudoscalar kaon, photon, nucleon, $\Lambda^*$, and vector kaon
fields, respectively. The $e_h$, $\kappa_h$, and $M_h$ denote the
electric charge, the anomalous magnetic moment, and the mass,
respectively, corresponding to the hadron $h$.  As for the
$K^*$-exchange contribution, we neglect the tensor coupling of the 
$K^{*}N\Lambda^*$ vertex, assuming its strength to be small.  

Above the threshold energy $E_\gamma \simeq 1900$ MeV, there are eight
nucleon resonances ($N^*$), as reported in Ref.~\cite{Yao:2006px}, up
to $\sqrt{s}\simeq 2200$ MeV: $P_{13}(1900)$, $F_{17}(1990)$, 
$F_{15}(2000)$, $D_{13}(2080)$, $S_{11}(2090)$, $P_{11}(2100)$,
$G_{17}(2190)$, $D_{15}(2200)$. Except for the $G_{17}$, their
confirmations are still poor (below two stars). Moreover,
we have little knowledge about the $N^*\to K\Lambda^*$ decay process in
comparison to other hyperons.  Hence, we exclude these resonance
contributions not to increase theoretical ambiguities for the moment.

We now discuss how to determine the coupling strength of the
$g_{KN\Lambda^*}$ briefly.  Although the $\Lambda^*$ does not decay 
into $\bar{K}N$ in free space, the value of the $g_{KN\Lambda^*}$ was
estimated within the potential model and chiral unitary model
($\chi$UM)~\cite{Jido:2002zk,Nacher:1998mi,Nam:2003ch}. In the
$\chi$UM, it was argued that the $\Lambda^*$ may consist of two 
individual poles~\cite{Jido:2002zk,Nacher:1998mi}. The pole positions
are $1398-74i$ MeV (lower one) and $1429-14i$ (higher one) in the
dimensional regularization scheme~\cite{Nam:2003ch}. The coupling
strengths to these poles were obtained from the residue of the
amplitude, resulting in $g_{KN\Lambda^*}=1.43$ for the lower pole and 
$g_{KN\Lambda^*}=2.52$ for the higher one in the dimensional
regularization\footnote{Here we take the absolute value of the
  $g_{KN\Lambda^*}$ computed in the $\chi$UM, since it is a complex
  number in general at the pole.}. When the dipole- and monopole-type
form factors are applied to the regularization of the loop integral,
one has $g_{KN\Lambda^*}=2.00$ and $3.64$, 
whereas $2.65$ and $3.39$, respectively~\cite{Nam:2003ch}.
Considering that the higher pole couples strongly to the $\bar{K}N$ state,
we take the average of those values of $g_{KN\Lambda^*}$, resulting in
$3.18$. This value is not much different from $1.5\sim3.0$, given in
Ref.~\cite{Williams:1991tw}.  The anomalous magnetic moment
$\kappa_{\Lambda^*}$ for the $\Lambda^*$ amounts to $0.44$ in
the SU(3) quark model~\cite{Williams:1991tw}.  From the $\chi$UM, it
was also estimated to be $0.24\sim0.45$~\cite{Jido:2002yz}. Therefore,
without loss of generality, we use $\kappa_{\Lambda^*}\approx0.4$ for
numerical calculations. The value  of $g_{K^*N\Lambda^*}$ is chosen to
be a free parameter, assuming that
$|g_{K^*N\Lambda^*}|{\le}g_{KN\Lambda^*}$. The $g_{\gamma K K^{*}}$
can be computed from experiments and reads $0.388\, \mathrm{GeV}^{-1}$
for the neutral decay and $0.254 \,\mathrm{GeV}^{-1}$ for the charged
decay~\cite{Yao:2006px}. 

Having performed a straightforward calculation by using the effective
Lagrangian given in Eq.~(\ref{eq:EL1405}), we obtain the invariant 
amplitudes for each diagram within the PS-coupling scheme as follows:    
\begin{eqnarray}
\label{eq:AMP1405}
i\mathcal{M}_s&=&-g_{KN\Lambda^*}
\bar{u}(p_2)\left[e_N\frac{\rlap{/}{k}_1+\rlap{/}{p}_1+M_N}{s-M^2_N} 
+\frac{e_Q\kappa_N}{2M_N}\frac{(\rlap{/}{p}_1+\rlap{/}{k}_1+M_N)}
{s-M^2_N}\rlap{/}{k}_1
\right]\rlap{/}{\epsilon}u(p_1)\times F_{\gamma NN}F_{KN\Lambda^*},
\nonumber\\
i\mathcal{M}_u&=&\frac{e_Q\kappa_{\Lambda^*}g_{KN\Lambda^*}}
{2M_{\Lambda^*}}
\bar{u}(p_2)\rlap{/}{\epsilon}\rlap{/}{k}_1
\frac{(\rlap{/}{p}_2-\rlap{/}{k}_1+M_{\Lambda^*})}
{u-M^2_{\Lambda^*}}u(p_1)
\times F_{KN\Lambda^*}F_{\gamma\Lambda^*\Lambda^*},
\nonumber\\
i\mathcal{M}^K_t&=&2e_Kg_{KN\Lambda^*}\bar{u}(p_2)
\frac{(k_2\cdot\epsilon)}{t-m^2_K}u(p_1)
\times F_{KN\Lambda^*}F_{\gamma KK},
\nonumber\\
i\mathcal{M}^{K^*}_t&=&ig_{\gamma KK^*}g_{K^*N\Lambda^*}
\bar{u}(p_2)\gamma_5\frac{\epsilon_{\mu\nu\sigma\rho}
k^\mu_1\epsilon^\nu k^\sigma_2\gamma^\rho}
{t-M^2_{K^*}}u(p_1)
\times F_{K^*N\Lambda^*}F_{\gamma KK^*},
\end{eqnarray}
where the Mandelstam variables are $s=(k_1+p_1)^2$, $u=(p_2-k_1)^2$,
and $t=(k_1-k_2)^2$, while $\epsilon_{\mu}$ the polarization 
vector for the incident photon.  The form factors for the
electromagnetic (EM) and hadronic vertices are given as: 
\begin{equation}
\label{eq:FF}
F_{\gamma\Phi\Phi(BB)}=\frac{\Lambda^2_\mathrm{EM}}{
  \Lambda^2_\mathrm{EM} +|{\bm k}_\gamma|^2},\,\,\,\, 
F_{\Phi BB}=\frac{\Lambda^2_h-M^2_\Phi}{\Lambda^2_h+|{\bm k}_\Phi|^2},
\end{equation}
where the subscripts $\Phi$ and $B$ stand for the mesonic and baryonic
particles involved, while $M$ and ${\bm k}$ are the on-shell mass and
the three momentum for the relevant particles.  The
$\Lambda_\mathrm{EM}$ and $\Lambda_h$ are the cutoff masses for the EM
and hadronic form factors, respectively.  In principle, the cutoff
mass corresponds to the inverse size of a hadron approximately. In
order to preserve the gauge invariance, since the $u$-  and
$K^*$-exchange channels are gauge-invariant by themselves, it is
enough to take $F_{\gamma KK}=F_{\gamma NN}$, similar sizes being
assumed for the proton ($\langle r^2\rangle^{1/2}_p\approx0.82$ fm)
and kaon ($\langle r^2\rangle^{1/2}_{K^+}\approx0.67$ fm), 
approximately. Moreover, we also set $F_{K^*N\Lambda^*} =
F_{KN\Lambda^*}$  and $F_{\gamma K^*K}=F_{\gamma KK}$ for simplicity.  

Assuming that the $\Lambda^*$ can be regarded as a molecular-type
$\bar{K}N$ bound state rather than a $uds$ color-singlet state, one
can infer that its size may be large in comparison to usual baryons 
such as the nucleon. In fact, we know from the phenomenological and
chiral potential model calculations that the absolute value of the EM
charge radius of $\Lambda^*$, $|\langle r^2\rangle^{1/2}_{\Lambda^*}|$
was estimated as $1.36\,\mathrm{fm}$~\cite{Yamazaki:2007cs} and   
$1.8\,\mathrm{fm}$~\cite{Hyodo:2007jq}, respectively, and 
its value was estimated to be $1.48\,
\mathrm{fm}$~\cite{Sekihara:2008qk} in the $\chi$UM. These values are
about two times larger than those of the typical baryons such as the
proton $\sim0.86\,\mathrm{fm}$.  Therefore, one may expect that the
cutoff mass for the form factor for the $\gamma\Lambda^*\Lambda^*$ can 
be smaller than usual baryons, corresponding to its larger spatial 
size. Thus, we choose the cutoff mass for the EM form factor to be
$650$ MeV for the all vertices as done for the $\Lambda(1520)$
photoproduction~\cite {Nam:2005uq}, except for the
$\gamma\Lambda^*\Lambda^*$ vertex in the $u$-channel, for which we
employ $\Lambda_\mathrm{EM}\approx300$ MeV, based on 
previous studies in various models.  Although the cutoff mass for the
hadronic form factors remain undetermined, we take it as the same as
that for the EM ones for brevity, $\Lambda_h\approx650$ MeV. We note 
that choosing a different cutoff mass only for the
$F_{\gamma\Lambda^*\Lambda^*}$ in the $u$-channel does not break the
gauge invariance, since the $u$-channel amplitude is gauge-invariant
by itself, as it contains only the magnetic coupling. All relevant
parameters used in the present work are listed in Table~\ref{table0}.   
\begin{table}[b]
\begin{tabular}{c|c|c|c|c}
$g_{KN\Lambda^*}$&$g_{K^*N\Lambda^*}$&$\kappa_{\Lambda^*}$&
$\Lambda_h$&$\Lambda_\mathrm{EM}$\\
\hline
$3.18$&$\pm3.18,0$&$0.4$&650 MeV&650(300) MeV\\
\end{tabular}
\caption{Relevant coupling strengths and anomalous magnetic moment, 
  and cutoff masses for the $\Lambda^*$ photoproduction. The
  $\Lambda_\mathrm{EM}=300$ MeV is only for the electromagnetic form
  factor $F_\mathrm{\gamma\Lambda^*\Lambda^*}$.} 
\label{table0}
\end{table}

\section{Numerical results\label{sec:3}}
\begin{figure}[ht]
\begin{tabular}{cc}
\includegraphics[width=7.5cm]{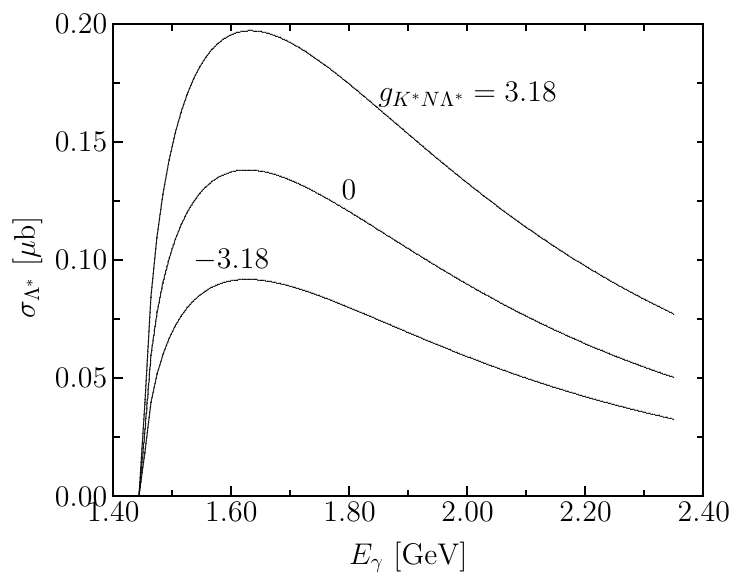}
\includegraphics[width=8cm]{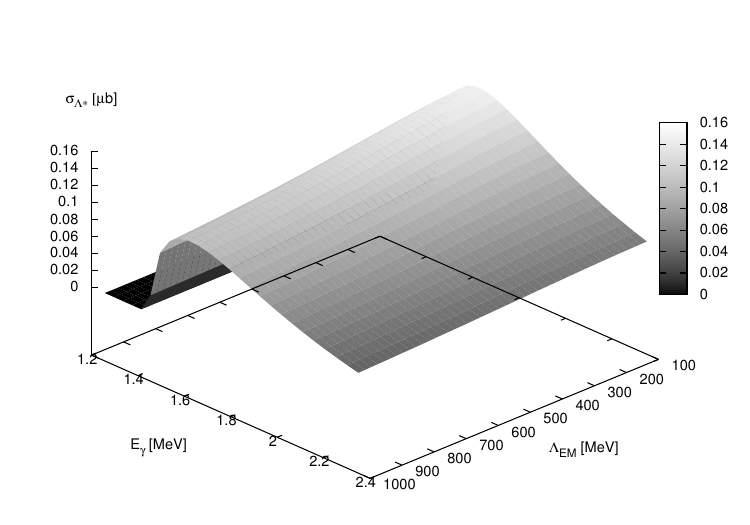}
\end{tabular}
\caption{In the left panel, the total cross section for the
$\Lambda^*$ photoproduction $\sigma_{\Lambda^*}$ for the different
coupling strengths of the $g_{K^*N\Lambda^*}\le|g_{KN\Lambda^*}|$  are
drawn as a function of the photon energy $E_\gamma$.  Here, we use the 
$\Lambda_\mathrm{EM}=300$ MeV for the
$F_{\gamma\Lambda^*\Lambda^*}$.  In the right panel, the total cross
section $\sigma_{\Lambda^*}$ is represented as a 
function of both the cutoff mass $\Lambda_\mathrm{EM}$ for the
$F_{\gamma\Lambda^*\Lambda^*}$ and the photon energy $E_\gamma$.}       
\label{fig1}
\end{figure}

We now provide the numerical results of the total and differential
cross sections, and the photon-beam asymmetry. First, we show those
for the total cross section as a function of the photon energy
$E_\gamma$ in the left panel of Fig.~\ref{fig1}.  We consider 
three different values of the $g_{K^*N\Lambda^*}$ ($0$ and
$\pm3.18$).  The results indicate that the total cross section
increases rapidly near the threshold, and then it starts to fall off
slowly as the photon energy increases. We note that the $s$-channel
contribution plays a main role in producing the present energy
dependence, whereas other contributions start to be effective beyond
about $E_\gamma=1.7$ GeV. The $K^*$-exchange contribution interferes
constructively ($+3.18$) and destructively ($3.18$) with other
contributions, as depicted in Fig~\ref{fig1}. The maximum value of the
magnitude of the total cross section turns out to be $0.1 \lesssim
\sigma_{\Lambda^*} \lesssim 0.2\,\mu \mathrm{b}$ near the threshold
depending on the value of $g_{K^*N\Lambda^*}$. We note that, however,
this tendency is rather different from that of the ground 
state $\Lambda(1116)$ photoproduction from the $\gamma p\to
K^+\Lambda(1116)$ reaction, in which the $t$-channel contribution 
is dominant to produce the appropriate energy
dependence~\cite{Janssen:2001wk,Ozaki:2007ka}.  In this sense, the 
$t$-channel dominance was argued that it also holds for the
$\Lambda^*$ photoproduction in Ref.~\cite{Lutz:2004sg} for
instance.  We note that, however, these different aspects are strongly 
dependent on the choice of the form factor schemes as will be
discussed in Sec.~\ref{sec:4}.   

We are now in a position to discuss the size effect of the 
$\Lambda^*$.  As mentioned in Sec.~\ref{sec:2}, we examine this effect
by introducing a phenomenological electromagnetic form factor
$F_{\gamma\Lambda^*\Lambda^*}$, varying the cutoff mass 
$\Lambda_\mathrm{EM}$ that reflects the size of the hyperon. In the
right panel of Fig.~\ref{fig1} we draw the total cross section for
$g_{K^*\Lambda^*\Lambda^*}=0$ as a function of the photon energy as
well as of the cutoff mass $\Lambda_\mathrm{EM}$ for the 
$F_{\gamma\Lambda^*\Lambda^*}$. We observe that as the size of the
$\Lambda^*$ decreases ($\Lambda_\mathrm{EM}$ increases), the magnitude
of the total cross section becomes larger around $E_\gamma\approx1.6$
GeV due to the enhancement of the $u$-channel contribution. For
instance, we can see from this figure that the maximum value of the
total cross section is about $0.12\,\mu\mathrm{b}$ for the 
$\Lambda_\mathrm{EM}=650$ MeV, whereas about $0.13\,\mu\mathrm{b}$ for
the $\Lambda_\mathrm{EM}=300$ MeV. Although we observe that the change
in the total cross section depends on the size of the $\Lambda^*$, it
is still small in comparison to theoretical ambiguities 
such as the $g_{K^*\Lambda^*\Lambda^* }$.  

In a recent experiment for the $\gamma p\to K^+\Lambda^*$ scattering 
process by the LEPS collaboration at SPring-8, it was shown that there
is a large difference in the $\Lambda^*/\Sigma(1385)$ production ratio
between the low ($1.5\lesssim E_\gamma\lesssim2.0$ GeV) and high energy
($2.0\lesssim E_\gamma\lesssim2.4$ GeV) regions~\cite{Niiyama:2008rt}.
This tendency may indicate that the energy dependence of the
$\sigma_{\Lambda^*}$ is qualitatively similar to our results as shown
in the left panel of Fig.~\ref{fig1}. Niyama et
al.~\cite{Niiyama:2008rt} also suggest that this interesting
tendency may be caused by either the novel internal structure of the
$\Lambda^*$ or the different production mechanisms.  From the present 
theoretical estimates, the $s$-channel dominance, being different from
the usual $t$-channel one shown in the ground state $\Lambda(1116)$
photoproduction, must be responsible for this large difference in the
ratio, whereas the size effect that encodes the novel internal
structure of the $\Lambda^*$ makes only small contribution.    

\begin{figure}[ht] \begin{tabular}{ccc}
\includegraphics[width=5.5cm]{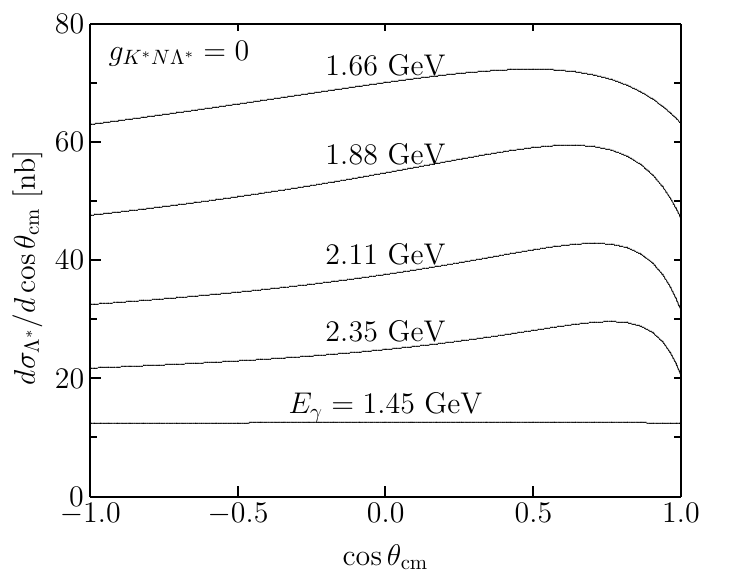}
\includegraphics[width=5.5cm]{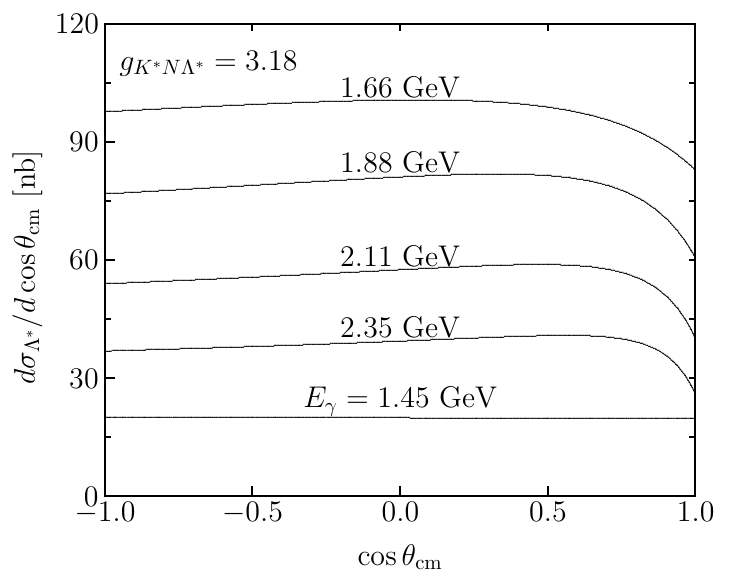}
\includegraphics[width=5.5cm]{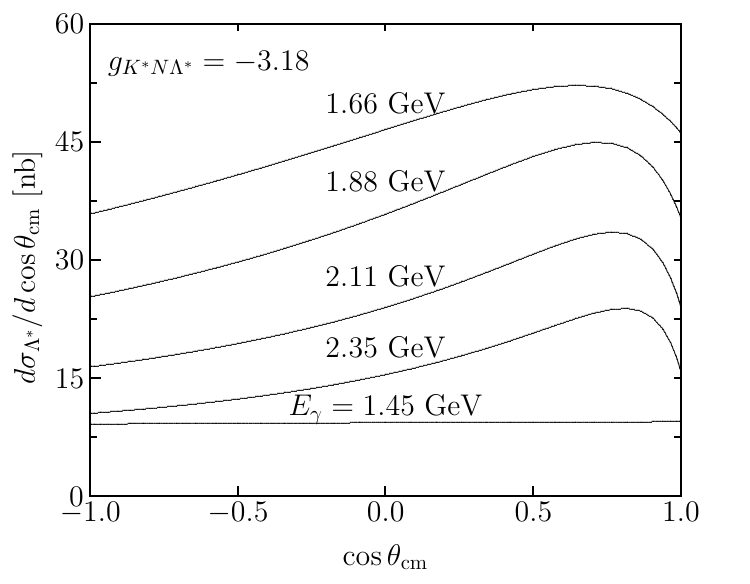}
\end{tabular}
\caption{Differential cross section
  $d\sigma_{\Lambda^*}/d\cos\theta_\mathrm{cm}$ as a function of 
  $\cos\theta_\mathrm{cm}$.  We also consider three different coupling
  strengths, $g_{K^*N\Lambda^*}=0$ (left), $3.18$ (middle) and $-3.18$
  (right), varying the photon energy $E_\gamma$.}       
\label{fig2}
\end{figure}
In Fig.~\ref{fig2}, we present the differential cross section as a
function of $\cos\theta_\mathrm{cm}$ in which the $\theta_\mathrm{cm}$
denotes the angle between the incident photon and the outgoing kaon in
the center of mass (cm) system. We also test the three different cases 
for the $g_{K^*N\Lambda^*}=0$ (left), $3.18$ (middle) and $-3.18$
(right), varying the photon energy $E_\gamma$ from 
$1.45$ GeV to $2.35$ GeV.  When the $K^*$-exchange contribution is
excluded as shown in the left panel of the figure, the angular
dependence shows the bump structure around
$\cos\theta_\mathrm{cm}\approx0.75$, and it becomes noticeable as the
photon energy increases. The main contributions to this bump 
are those of the $s$-channel and the $t$-channel, which enhance
the differential cross section with the bump in the forward region.
Note that the $u$-channel contribution turns out to be negligible.   

For a finite $g_{K^*N\Lambda^*}$ as in the middle 
($g_{K^*N\Lambda^*}=3.18$) and right ($g_{K^*N\Lambda^*}=-3.18$)
panels, the angular dependence of the differential cross section is 
changed obviously, since the $K^*$-exchange contributes mildly to the 
backward region.  Due to the distinctive interference pattern
between the $K^*$-exchange contribution and others, especially to the
$t$-channel, the bump in the forward region gets diminished 
($g_{K^*N\Lambda^*}=3.18$) or enhanced ($g_{K^*N\Lambda^*}=-3.18$) in
comparison to the case without the $K^*$-exchange.  If we take a larger
(smaller) value for the $\Lambda_\mathrm{EM}$ to test the size effect
of the $\Lambda^*$, it turns out that only the magnitude of the curves
become slightly smaller (larger), whereas the angular dependence
remains almost the same, as expected from the results shown in the
right panel of Fig.~\ref{fig1}.  

In Ref.~\cite{Williams:1991tw}, the differential cross section
was computed for the $g_{KN\Lambda^*}=1.5$ and $3.0$ with the
nucleon-resonance contributions from $N^*(1650)$ and $N^*(1710)$. It
showed a small forward-scattering enhancement, being similar to ours
qualitatively, but the order of magnitude of the 
differential cross section in their work is about three or four times  
larger than ours when $g_{KN\Lambda^*}=3.0$. From the experimental
data~\cite{Niiyama:2008rt}, the differential cross section was
estimated to be about $0.4\,\mu$b for $1.5\lesssim
E_\gamma\lesssim2.0$ GeV and
$0.8\lesssim\cos\theta_\mathrm{cm}\lesssim1.0$. This value is about
four and two times larger than ours and that of
Ref.~\cite{Nacher:1998mi}, respectively, but very similar to that of  
Ref.~\cite{Williams:1991tw}, although all of them are in a similar
order.  

Now we define the photon-beam asymmetry, which is an important
physical observable in the photoproduction, as follows:  
\begin{equation}
\label{eq:BA}
\Sigma_{\Lambda^*}=\frac{d\sigma_{\Lambda^*\perp}
-d\sigma_{\Lambda^*\parallel}}{d\sigma_{\Lambda^*\perp}
+d\sigma_{\Lambda^*\parallel}},
\end{equation}
where the subscript $\perp$ ($\parallel$) indicates that the incident
photon is polarized transversely (longitudinally) to the reaction
plane. We note that by definition the $\Sigma_{\Lambda^*}$ becomes
positive for the magnetic photon-hadron coupling contribution, whereas
the negative for the electric one. We observe that the $u$-channel and
$K^*$- exchange contributions give finite positive values for the
$\Sigma_{\Lambda^*}$. On the contrary, it becomes $-1$ for the
$K$-exchange, and negligible for the $s$-channel contribution.  

\begin{figure}[b]
\begin{tabular}{ccc}
\includegraphics[width=5.5cm]{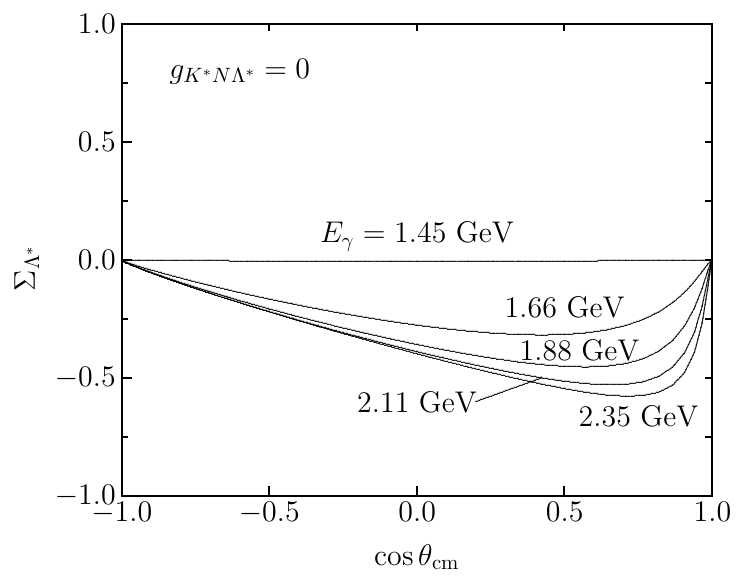}
\includegraphics[width=5.5cm]{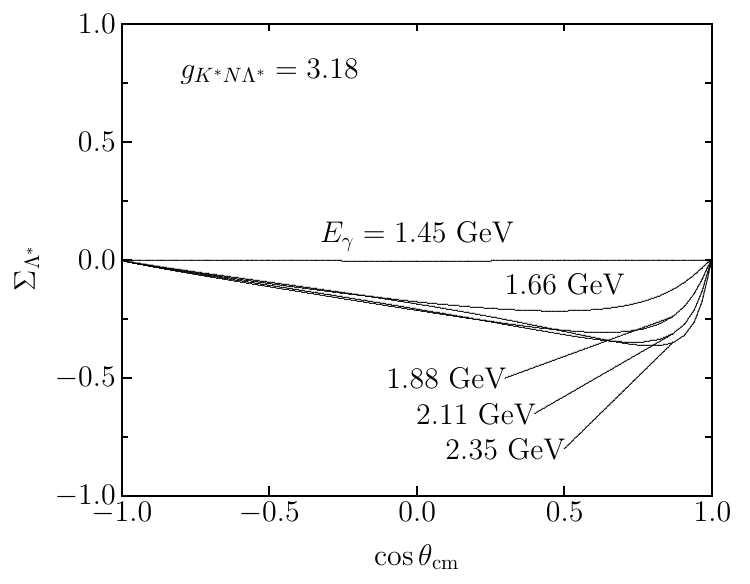}
\includegraphics[width=5.5cm]{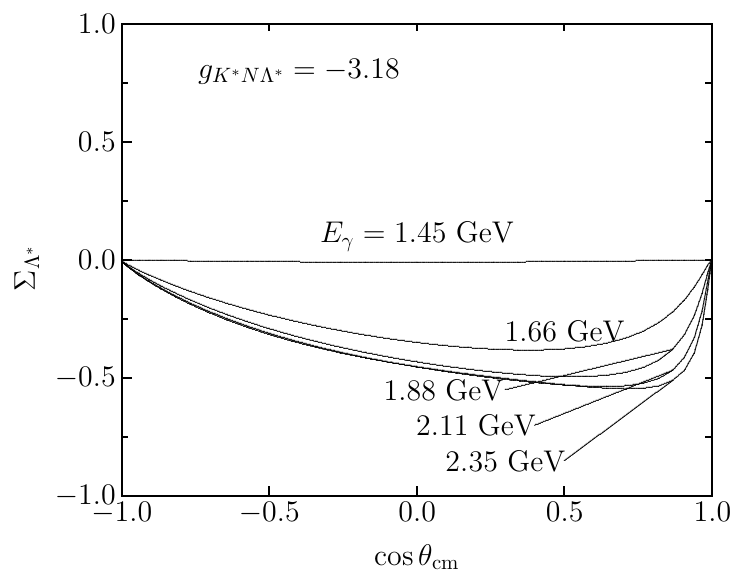}
\end{tabular}
\caption{Photon-beam asymmetry as a function of
  $\cos\theta_\mathrm{cm}$.  
We also consider the three different coupling strengths, i.e. 
$g_{K^*N\Lambda^*}=0$ (left), $3.18$ (middle) and $-3.18$ (right)
varying the photon energy $E_\gamma$.}       
\label{fig3}
\end{figure}
In Fig.~\ref{fig3}, we show the numerical results for the
$\Sigma_{\Lambda^*}$ as a function of $\cos\theta_\mathrm{cm}$ varying
the strength of the $g_{K^*N\Lambda^*}=0,\,3.18,\,-3.18$. We also
alter the photon energy,
$E_\gamma=1.45\,\mathrm{GeV}\sim2.35\,\mathrm{GeV}$. In  
Fig.~\ref{fig3}, we observe that the $\Sigma_{\Lambda^*}$ becomes
negative for all the cases due to the strong contribution of the electric
coupling from the $K$-exchange.  In the left panel of Fig.~\ref{fig3},
we draw the results with $g_{K^*N\Lambda^*}=0$. The
$\Sigma_{\Lambda^*}$ becomes tilted negatively in the forward 
region ($0.5\lesssim\cos\theta_\mathrm{cm}$) because of
the destructive interference between the $u$-channel and
$K$-exchange contributions.  Even though we include the
magnetic-coupling contributions such as the $K^*$-exchange, the
qualitative shapes of the curves for the beam 
asymmetry remain almost unchanged as shown in the middle
($g_{K^*N\Lambda^*}=3.18$) and right ($g_{K^*N\Lambda^*}=-3.18$)
panels.  It indicates that only small interferences with the
$K^*$-exchange contribution appear.  The photon-energy dependence of
the $\Sigma_{\Lambda^*}$ shows the mild enhancement of the bump around
$0.5\lesssim\cos\theta_\mathrm{cm}\lesssim1.0$, because of the
$u$-channel contribution.  We also verified that the size effect from
the $\Lambda^*$ is hard to be seen in the photon-beam asymmetry.
\section{Discussions on theoretical ambiguities\label{sec:4}}
In this Section, we want to discuss theoretical ambiguities,
which may make significant effects on the present results given in 
Sec.~\ref{sec:3}.  We can consider the three theoretical ambiguities
that are most critical in the present work:
\begin{itemize}
\item The coupling strengths of the $g_{KN\Lambda^*}$ and the
  $g_{K^*N\Lambda^*}$, 
\item The form factor scheme,
\item The resonance contributions.
\end{itemize} 

As explained in Sec.~\ref{sec:3}, we have used the
$g_{KN\Lambda^*}=3.18$, which is derived by averaging the possible
values of the $g_{KN\Lambda^*}$ computed from the $\chi$UM with
different regularization schemes, considering only the higher pole for
the $\Lambda^*$ which couples to the $\bar{K}N$ state strongly in 
comparison to the lower one. If we take the lower pole additionally,
the coupling strength becomes smaller, $g_{KN\Lambda^*}=2.6$.  With
this value, the maximum value of the total cross section gets lowered
by about $30\sim40\%$. The phenomenological
estimations~\cite{Williams:1991tw}, $g_{KN\Lambda^*}=1.5\sim3.0$ can 
also give larger uncertainties but does not change the results much
qualitatively, in particular, the energy dependence. 

As for the $g_{K^*N\Lambda^*}$, the overall shapes of the total cross
sections and the photon-beam asymmetry do not depend much on
it within our present choice of the parameter range. On the contrary,
the differential cross section (angular dependence) is affected by
different values of the $g_{K^*N\Lambda^*}$ due to the complicated
interference pattern between the $K^*$-exchange contribution and
others. The other $\Lambda$ hyperons being considered, the theoretical 
estimations of the ratio $|g_{K^*N\Lambda(1520)}/g_{KN\Lambda(1520)}|$
is very small $\sim0.1$~\cite{Hyodo:2006uw}.  On the other hand, the
Nijmegen potential suggests $|g_{K^*N\Lambda(1116)} /
g_{KN\Lambda(1116)}| \approx5$~\cite{Stoks:1999bz}. Thus, it is rather
difficult to choose a reasonable value for $g_{KN\Lambda^*}$ from the
phenomenological point of view. We also note that from the pure
duality consideration as in Ref.~\cite{Williams:1991tw}, the
$K^*$-exchange contribution can be ignored. If this is the case, we
can assume that the $g_{K^*N\Lambda^*}$ is not large. 

We can also choose the different schemes for the form factors as in
Refs.~\cite{Ohta:1989ji,Haberzettl:1998eq,Davidson:2001rk}:  
\begin{equation}
\label{eq:FDFF}
F(x)=\frac{\Lambda^4}{\Lambda^4+(x-m^2_x)^2},
\end{equation}
where $x$ and $m_x$ stand for the Mandelstem variable and the
intermediate hadron with the off-shell momentum squared $q^2=x$,
respectively.  The $\Lambda$ is the four-dimensional cutoff mass. The 
detailed explanation for its usage can be found in
Refs.~\cite{Nam:2004xt,Ozaki:2007ka}. This form factor preserves the
Ward-Takahashi identity, and one of its typical features lies in the
fact that it suppresses the $s$- and $u$-channel contributions,
leading to the $t$-channel dominance, when it is applied to the
spin-1/2 baryon photoproduction~\cite{Nam:2004xt,Ozaki:2007ka}.  In
this case, the contributions from the $s$- and $u$-channels become
small or negligible. Consequently, the angular dependence computed in
this scheme usually shows a strong enhancement in the forward
direction due to the $t$-channel.  Especially, the $t$-channel
contribution has been argued as the dominant one to reproduce the
experimental data of the ground state $\Lambda(1116)$
photoproduction~\cite{Janssen:2001wk, Hyodo:2004vt, Ozaki:2007ka}. 

\begin{figure}[t]
\begin{tabular}{cc}
\includegraphics[width=7.5cm]{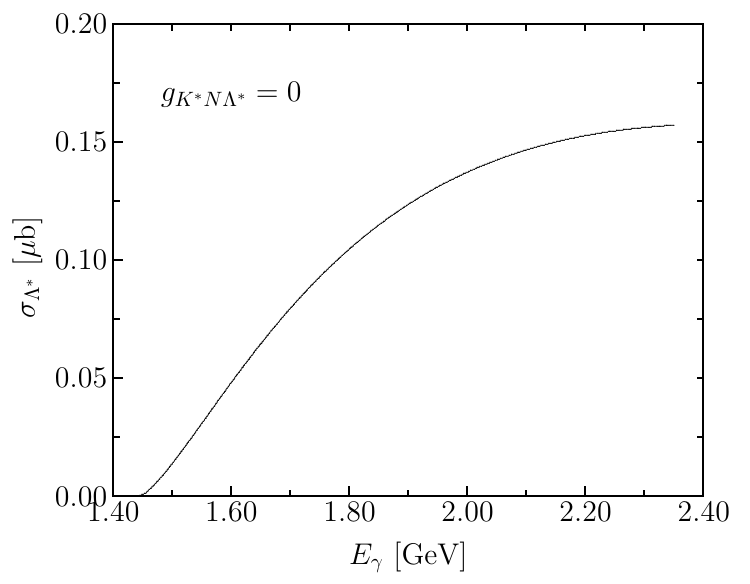}
\includegraphics[width=7.5cm]{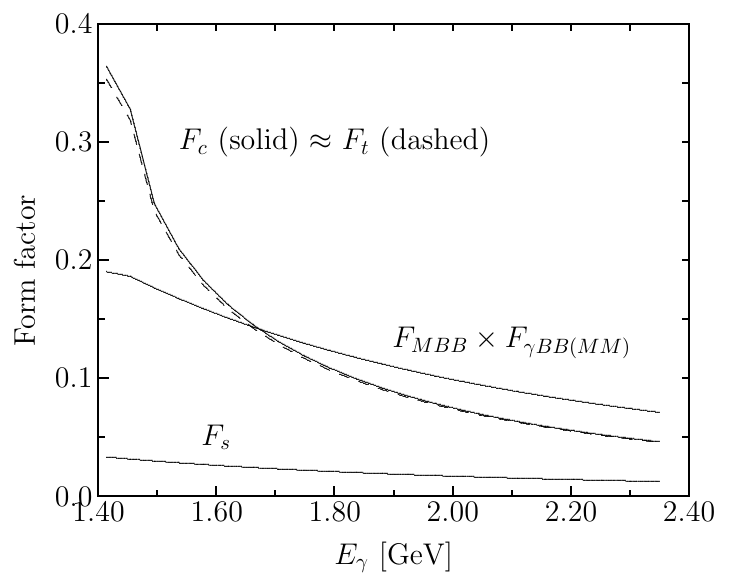}
\end{tabular}
\caption{In the left panel, we draw the total cross section as a
  function of the photon energy $E_\gamma$ with the four dimensional
  gauge-invariant form factor given in 
  Eq.~(\ref{eq:FDFF}).  In the right panel, we depict the form factors
  given in Eqs.~(\ref{eq:FF}) and (\ref{eq:FDFF}) as a function of
  the photon energy $E_\gamma$, respectively.}        
\label{fig5}
\end{figure}

We draw the energy dependence in the left panel of Fig.~\ref{fig5} for
the $\Lambda^*$ photoproduction using the form factor given in
Eq.~(\ref{eq:FDFF}) with the cutoff mass $\Lambda=700$ MeV.  As shown 
in Fig.~\ref{fig5}, the total cross section increases slowly, starting
from the threshold, and then becomes almost saturated in the vicinity
of $E_\gamma\approx2.2$ GeV. This tendency is very different from that 
shown in the left panel of Fig.~\ref{fig1} for which the form factors
of Eq.~(\ref{eq:FF}) are used.  We also have verified that the
$K^*$-exchange brings out only a small deviation from the
curve given in Fig.~\ref{fig5}.  

We now look carefully into these form factor
schemes. In the right panel of Fig.~\ref{fig5} we depict the various
form factors as a function of the photon energy which are applied to
the invariant amplitudes in a gauge-invariant manner for the
$\Lambda^*$ photoproduction as follows:  
\begin{eqnarray}
\label{eq:}
i\mathcal{M}^{\mathrm{Eq.~(3)}}_\mathrm{total}
&=&F_{\gamma BB(MM)}F_{MBB}
\left[i\left(\mathcal{M}^E_s+\mathcal{M}^M_s \right)+i\mathcal{M}_t
+i\mathcal{M}^M_u\right],
\nonumber\\
i\mathcal{M}^{\mathrm{Eq.~(5)}}_\mathrm{total}&=&
\left[i\left(F_c\,\mathcal{M}^E_s+F_s\,\mathcal{M}^M_s \right)
+iF_c\,\mathcal{M}_t+iF_u\,\mathcal{M}^M_u\right],
\end{eqnarray}
where we ignore the $K^*$-exchange contribution for simplicity.  The  
$\mathcal{M}^E_s$ means the electric part of the $s$-channel
amplitude,  whereas $\mathcal{M}^M_s$ that of the magnetic
one. Following 
Refs.~\cite{Ohta:1989ji,Haberzettl:1998eq,Davidson:2001rk}, we can
define the $F_c$ as $F_s+F_t-F_sF_t$ so that it may satisfy the
on-shell condition, $F(0)=1$. From Fig.~\ref{fig5}, we see that the
$F_c$ is almost the same as $F_t$, while the $F_s$ is much smaller
than them. The $F_u$ is also negligible in comparison to the
$F_c$. The $F_{\gamma BB(MM)}F_{MBB}$ lies between $F_c$ and
$F_s$. Hence, in the form factor scheme of Eq.~(\ref{eq:FDFF}), the
differences between the channels become clear: The $t$-channel is
effective much more than others. On the contrary, the form factor
given in  Eq.~(\ref{eq:FF}) suppresses all the channels simultaneously
as the photon energy increases. Thus, the $s$-channel contribution can
dominate the process near the threshold.  

Finally, we may expect possible contributions from the nucleon,
$\Delta$ and hyperon resonances, which have been excluded in the
present work.  For instance, Oh {\it et al.}~\cite{Oh:2007jd} studied
the $\Sigma(1385)$ photoproduction in a similar method to the present
work, but with various resonance contributions, whose relevant
coupling strengths to the $\Sigma(1385)$ were derived theoretically,
the SU(3) relation being considered.  From their analysis, it turned
out  that the resonance contribution plays an important role for the
region from the threshold to $E_\gamma\approx2$ GeV. If this is the
case also for the $\Lambda^*$ photoproduction, although they are
different in the spin and parity, and information on the $\Lambda^*$
related to the resonances is very poor in comparison to the
$\Sigma(1385)$, the present results can be altered to a large extent.  
\section{Summary and Conclusion}
In the present work, we have investigated the
$\Lambda(1405,1/2^-)\equiv\Lambda^*$ photoproduction, employing the  
effective Lagrangian approach in the Born approximation. We took into
account the minimal contributions, the $s$- and $u$-channels, and the
pseudoscalar $K$- and vector $K^*$-exchange contributions in the
$t$-channel, without nucleon- and hyperon-resonance contributions 
such that the analysis of the $\Lambda(1405,1/2^-)$ photoproduction
mechanism can be easily achieved.  All the necessary parameters were
determined from the possible theoretical (the $\chi$ unitary model)
and experimental (the $\Lambda(1520)$ photoproduction)
results.  The phenomenological electromagnetic and  hadronic form
factors were introduced with the gauge invariance preserved.  

Assuming the molecular-type $\bar{K}N$ bound state for the
$\Lambda^*$ rather than the $uds$ color-singlet one, we examined the 
size effect of the $\Lambda^*$ by changing the cutoff mass for the
electromagnetic form factor sitting on the $\gamma\Lambda^*\Lambda^*$
interacting vertex. Choosing this cutoff mass
$\Lambda_\mathrm{EM}\approx300$ MeV, which may correspond to a larger 
spatial size of the $\Lambda^*$ and is about a half of the other
cutoff mass $\Lambda_h\approx650$ MeV, we observed that, depending on
the choice of the $K^*N\Lambda^*$ coupling strength, the total cross 
section turns out to be $0.1 \lesssim \sigma_{\Lambda^*} \lesssim
0.2\, \mu \mathrm{b}$ near the threshold and decreases slowly beyond 
$E_\gamma\approx1.6$ GeV, showing the $s$-channel dominance. The
angular dependence shows a mild enhancement in the forward direction
due to the $K$-exchange in the $t$-channel. By the same reason, the
photon-beam asymmetry resulted in the electric-coupling dominance
(negative photon-beam asymmetry) for all angle regions.

Concerning the size effect of the $\Lambda^*$, had we considered
its larger size, the bump at $E_\gamma\approx1.6$ GeV shown in the
total cross section would have increased slightly and smoothly,
because of the enhancement of the $u$-channel contribution near the 
threshold. However, we note that the size effect is hard to be seen in
all the physical observables computed in the present
work. Consequently, we note that the enhancement of the total cross
section near the threshold, as reported by the LEPS experiment, can
depend much on the different production mechanisms, as shown in
Sec.~\ref{sec:3}, not on the novel internal structure of the
$\Lambda^*$.    

Finally, we discussed the theoretical ambiguities which can make
effects on the present results given in Sec.~\ref{sec:3}.  Among them,
while the coupling strengths made little change the physical
observables, it turned out that the form factor schemes are of great
significance in the present results.  The contributions from higher
resonances may play also an important role in describing the mechanism
of the $\Lambda^*$ photoproduction.  The corresponding works are under
progress.

\section*{Acknowledgments}
The authors would like to thank J.~K.~Ahn, T.~Nakano, D.~Jido,
M.~Niiyama, and H.~Kohri for fruitful discussions. The work of
H.C.K. is supported by Basic Science Research Program through the
National Research Foundation of Korea (NRF) funded by the Ministry of
Education, Science and Technology (grant number: 2009-0089525).

 
\end{document}